\documentclass[prb,showpacs,byrevtex,floatfix,twocolumn]{revtex4}
\usepackage{graphicx}
\usepackage{dcolumn}
\usepackage{bm}

\begin{document}

\bibliographystyle{apsrev}

\preprint{Draft - not for distribution}

\title[FGT]{\boldmath Sum rules and energy scales in the
  high-temperature superconductor YBa$_2$Cu$_3$O$_{6+x}$ \unboldmath}

\author{C. C. Homes}
\email{homes@bnl.gov}%
\author{S. V. Dordevic}
\affiliation{Department of Physics, Brookhaven National Laboratory,
  Upton, New York 11973}
\author{D. A. Bonn}
\author{Ruixing Liang}
\author{W. N. Hardy}
\affiliation{Department of Physics and Astronomy, University of British
Columbia, Vancouver, British Columbia, Canada V6T 1Z1}
\date{\today}

%
%
\begin{abstract}
The Ferrell-Glover-Tinkham (FGT) sum rule has been applied to the temperature
dependence of the in-plane optical conductivity of optimally-doped
YBa$_2$Cu$_3$O$_{6.95}$ and underdoped YBa$_2$Cu$_3$O$_{6.60}$.  Within the
accuracy of the experiment, the sum rule is obeyed in both materials. However,
the energy scale $\omega_c$ required to recover the full strength of the
superfluid $\rho_s$ in the two materials is dramatically different;
$\omega_c\simeq 800$~cm$^{-1}$ in the optimally doped system (close to twice
the maximum of the superconducting gap, $2\Delta_0$), but $\omega_c\gtrsim
5000$~cm$^{-1}$ in the underdoped system.
In both materials, the normal-state scattering rate close to the critical
temperature is small, $\Gamma < 2\Delta_0$, so that the materials are not in
the dirty limit and the relevant energy scale for $\rho_s$ in a BCS material
should be twice the energy gap.
The FGT sum rule in the optimally-doped material suggests that the majority of
the spectral weight of the condensate comes from energies below $2\Delta_0$,
which is consistent with a BCS material in which the condensate originates from
a Fermi liquid normal state.  In the underdoped material the larger energy
scale may be a result of the non-Fermi liquid nature of the normal state.  The
dramatically different energy scales suggest that the nature of the normal
state creates specific conditions for observing the different aspects of what
is presumably a central mechanism for superconductivity in these materials.
\end{abstract}

%
%
%

\maketitle


%
%
\section{Introduction}
Sum rules and conservation laws play an important role in physics.  In
spectroscopy, the conductivity sum rule is particularly useful and is an
expression of the conservation of charge.\cite{smith}  In metallic systems, the
conductivity sum rule usually yields the classical plasma frequency or the
effective number of carriers. In superconductors, below the critical
temperature $T_c$ some fraction of the carriers collapse into the
$\delta(\omega)$ function at zero frequency that determines the London
penetration depth $\lambda_L$, with a commensurate loss of spectral weight from
low frequencies (below twice the superconducting energy gap,$2\Delta$). This
shift in spectral weight may be quantified by the application of the
conductivity sum rule to the normal and superconducting states, as discussed by
Ferrell, Glover and Tinkham (the FGT sum rule),\cite{ferrell58,tinkham59} which
is used to estimate the strength of the superconducting condensate $\rho_s =
c^2/\lambda_L^2$. The theory of superconductivity described by Bardeen, Cooper
and Schrieffer (BCS) holds that while the kinetic energy of the superconducting
state is greater than that of the normal state,\cite{bcs,tinkham} this increase
is compensated by the reduction in potential energy which drives the transition
(the net reduction of energy is simply the condensation energy).  However, it
has been proposed that in certain hole-doped materials the superconductivity
could arise from a lowering of the kinetic rather than the potential
energy.\cite{hirsch92} In such a system non-local transfers of spectral weight
would result in the apparent violation of the FGT sum rule, which would yield a
value for the strength of $\rho_s$ that would be too small ($\lambda_L$ would
be too large).\cite{hirsch92,hirsch00a,hirsch00b} Similar models in the cuprate
materials presume either strong coupling,\cite{norman00,haslinger02} or that
the normal state is not a Fermi liquid and that superconductivity is driven
either by the recovery of frustrated kinetic energy when pairs are
formed,\cite{lee99,anderson00} by lowering the in-plane zero-point kinetic
energy,\cite{emery00} or by the condensation of preformed
pairs.\cite{alexandrov94}

%
%
Experimental results along the poorly-conducting interplane ({\it c}-axis)
direction in several different cuprate materials support the view that the
kinetic energy is reduced below the superconducting transition.\cite{klein99,
basov99,basov01}  In some materials, the low-frequency {\it c}-axis spectral
weight accounts for only half of the strength of the condensate. However, this
violation of the FGT sum rule appears to be restricted to the underdoped
materials which display a pseudogap in the conductivity.\cite{timusk99} The
dramatically lower value of the strength of the condensate along the {\it c}
axis makes it easier to observe kinetic energy contributions.
%
%
In comparison, the much larger value of the condensate in the copper-oxygen
planes makes it much more difficult to observe changes due to the kinetic
energy based on optical sum rules.\cite{norman02,vdmprivate} Recently,
high-precision measurements in the near-infrared and visible region have
reported small changes in the in-plane spectral weight associated with the
onset of superconductivity, supporting the argument that changes in the kinetic
energy are indeed occurring.\cite{molegraaf02,syro01,syro02} While the
relatively small changes of the in-plane spectral weight make it difficult to
make statements about the kinetic energy, it is nonetheless a strong motivation
to examine the evolution of the spectral weight in detail to see if there are
unexpected signatures of an unconventional mechanism for the superconductivity
in this class of materials.


%
%
In this paper we examine the changes of the in-plane spectral weight and the
evolution of the superconducting condensate in the optimally doped and
underdoped detwinned YBa$_2$Cu$_3$O$_{6+x}$ single crystals for light polarized
perpendicular to the copper-oxygen chains, along the {\it a} axis. The BCS
model requires that the spectral weight of the condensate be fully formed at
energies comparable to the energy gap ($\omega_c \simeq 2\Delta$), with no
subsequent violation of the FGT sum rule. This is precisely what is observed
for the optimally-doped material, within the limits of experimental accuracy
for the sum rules, which is estimated to be about 5\%. However, in the
underdoped material only 80\% of the spectral weight of the condensate has been
recovered at energies comparable to $2\Delta$; the FGT sum rule must be
extended to considerably higher frequencies to recover the remaining spectral
weight ($\omega_c\gtrsim 0.6$~eV).  For $T\gtrsim T_c$, the normal-state
scattering rate is determined to be small, so this shift in spectral weight can
not be attributed to dirty-limit effects in response to impurities. However,
the nature of the normal state is dramatically different in the optimally-doped
material, which is reminiscent of a Fermi liquid, and the underdoped material,
which develops a pseudogap and displays non-Fermi liquid behavior.  The
dramatically different energy scales required to recover the full value of
$\rho_s$ suggest that these aspects of the superconductivity are related to the
normal state properties from which it emerges.

%
%
\section{Experiment and Sample Preparation}
Details of the growth and characterization of the mechanically-detwinned
YBa$_2$Cu$_3$O$_{6+x}$ crystals have been previously described in
detail\cite{liang92,schleger91} and will be discussed only briefly.  The
crystal had a small amount of Ni deliberately introduced, Cu$_{1-x}$Ni$_x$,
where $x=0.0075$.  Such a small concentration of Ni results in a critical
temperature which is slightly lower ($\sim 2$~K) than the pure materials, with
a somewhat broader transition .  The same detwinned crystal has been carefully
annealed to produce two different oxygen concentrations, $x=0.95$ ($T_c \simeq
91$~K) and $0.60$ ($T_c \simeq 57$~K).  The reflectance for light polarized
along the {\it a} axis (perpendicular to the CuO chains, therefore probing only
the CuO$_2$ planes) has been measured at a variety of temperatures over a wide
frequency range ($\approx 40$ to 9000~cm$^{-1}$) using an overfilling
technique;\cite{homes93a} this reflectance has been extended to very high
frequency ($3.5\times 10^5$~cm$^{-1}$) using the data of Basov {\it et
al.}\cite{basov95,basov96} and Romberg {\it et al.}\cite{romberg90}  The
absolute value of the reflectance is estimated to be accurate to within 0.2\% .
The optical properties are calculated from a Kramers-Kronig analysis of the
reflectance. The optical conductivity, which dealt with the effects of Ni
doping on the CuO chains and the reduction of the in-plane anisotropy, has been
previously reported.\cite{homes99}  The presence of Ni in such small
concentrations was not observed to have any effect on the conductivity of the
CuO$_2$ planes in either the normal or superconducting states.

%
%
\section{Results and Discussion}
%
%
\subsection{Optical sum rules}
Optical sum rules comprise a powerful set of tools to study and characterize
the lattice vibrations and electronic properties of solids.  The spectral
weight may be estimated by a partial sum rule of the
conductivity\cite{smith,neff}
\begin{equation}
   N(\omega_c)= {120\over \pi}\int_0^{\omega_c} \sigma_1(\omega)\,d\omega \,
   {\buildrel \omega_c \to \infty \over \longrightarrow} \, \omega_p^2
\end{equation}
%
%
where $\omega_p^2=4\pi ne^2/m_0$ is the classical plasma frequency, $n$ is the
carrier concentration, and $m_0$ is the bare optical mass. In the absence of
bound excitations, this expression is exact in the limit of $\omega_c
\rightarrow \infty$. However, any realistic experiment involves choosing a
low-frequency cut-off $\omega_c$.
When applied to the Drude model
\begin{equation}
  \tilde\epsilon(\omega) = \epsilon_\infty - {{\omega_{p}^2} \over {\omega
  (\omega + i\Gamma)}},
\label{eq:drude}
\end{equation}
where $\epsilon_\infty$ is the contribution from the ionic cores, and
$\Gamma=1/\tau$ is the scattering rate, the conductivity sum rule indicates
that 90\% of the spectral weight is recovered for $\omega_c \sim 6\Gamma$. For
even modest choices of $\Gamma$, $\omega_c$ can be quite large ($\simeq
\omega_p$). This places some useful constraints on the confidence limits for
the conductivity sum rule in the normal state.

%
%
The conductivity in the superconducting state for any polarization {\bf r} has
two components\cite{tinkham}
\begin{equation}
  \sigma_{1,\mathbf{r}}^{SC}(\omega) = {\pi\over 120}\rho_{s,\mathbf{r}}\delta(\omega) +
  \sigma_{1,\mathbf{r}}^{reg}(\omega).
\end{equation}
The first part is associated with the superconducting $\delta(\omega)$ function
at zero frequency, where $\rho_{s,\mathbf{r}}$ is the superfluid stiffness, or
strength of the superconducting order.\cite{emery95a} This is often expressed
as the square of a plasma frequency
$\omega^2_{pS}=4\pi{n_s}e^2/m_\mathbf{r}^\ast$, where $n_s$ is the density of
superconducting electrons, and $m^\ast_\mathbf{r}$ is the effective mass
tensor. The second component $\sigma_{1,\mathbf{r}}^{reg}(\omega)$ is referred
to as the ``regular'' component for $\omega > 0$ and is associated with the
unpaired charge carriers.

%
%
\begin{figure}[t]
\vspace*{-0.4cm}%
%
%
\centerline{\includegraphics[width=3.8in]{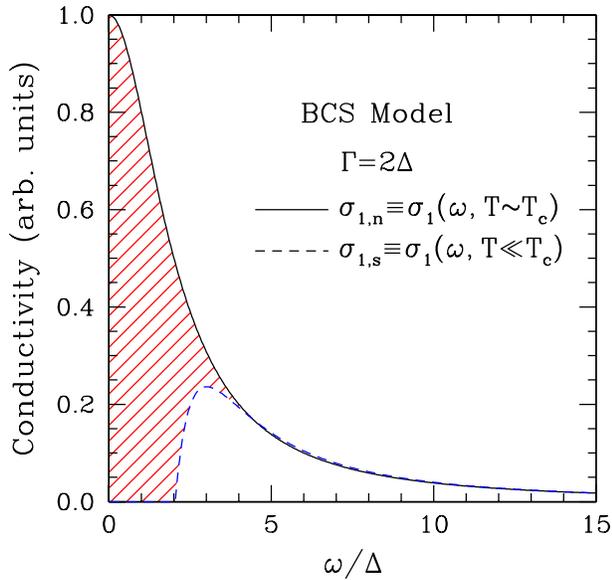}}%
%
%
%
\vspace*{-1.2cm}%
\caption{The real part of the optical conductivity calculated for a BCS model
for a normal-state scattering rate of $\Gamma=2\Delta$.
The conductivity in the normal state $\sigma_1(\omega, T\gtrsim T_c)$ is shown
by the solid line (normalized to unity), while the conductivity in the
superconducting state $\sigma_1(\omega,T\ll T_c)$ is shown by the dashed line.
For $T\ll T_c$ the superconducting gap $2\Delta$ is fully formed and there is
no absorption below this energy.  The hatched area illustrates the spectral
weight that has collapsed into the superconducting $\delta(\omega)$ function at
the
origin.}%
\label{fig:bcs}
\end{figure}

\begin{figure}[t]
\vspace*{-0.4cm}%
%
%
\centerline{\includegraphics[width=3.8in]{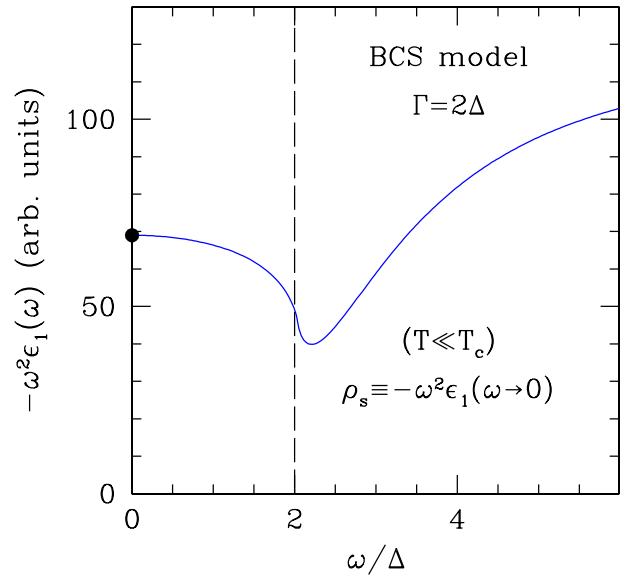}}%
%
%
%
\vspace*{-1.2cm}%
\caption{The value of $-\omega^2\epsilon_1(\omega)$ calculated from the BCS
model for $T\ll T_c$ for the normal-state scattering rate $\Gamma = 2\Delta$
(solid line); $\rho_s=-\omega^2\epsilon_1(\omega)$ in the limit of $\omega
\rightarrow 0$.  Note that there is also a slight minima near $2\Delta$
(long dashed line).}%
\label{fig:sigma2}
\end{figure}

A variation of the conductivity sum rule in a superconductor is to study the
amount of spectral weight that collapses into the superconducting
$\delta(\omega)$ function at the origin below the critical
temperature.\cite{tinkham} This scenario is represented in Fig.~\ref{fig:bcs},
which shows the normalized conductivity for a BCS {\it s}-wave model for an
arbitrary purity level\cite{zimmerman91} where the scattering rate in the
normal state is chosen as $\Gamma=2\Delta$ ($2\Delta$ is the full gap value for
$T\ll T_c$). The solid line shows the real part of the optical conductivity
$\sigma_1(\omega)$ in the normal state for $T\gtrsim T_c$, while the dashed
line is the calculated value for $\sigma_1(\omega)$ in the superconducting
state for $T\ll T_c$.  For $T\ll T_c$ the gap is fully formed, and there is no
conductivity for $\omega < 2\Delta$, above which the onset of absorption
occurs.
%
%
The missing spectral weight represented by the hatched area represents the
strength of the condensate $\omega_{pS}^2$.  This area may be estimated by the
FGT sum rule\cite{ferrell58,tinkham59}
%
%
\begin{equation}
  \omega_{pS}^2={120\over\pi} \int_{0+}^{\omega_c} \left[
  \sigma_{1,n}(\omega) - \sigma_{1,s}(\omega) \right]\,d\omega .
\end{equation}
where $\sigma_{1,n}(\omega) \equiv \sigma_{1}(\omega,T\gtrsim T_c)$ and
$\sigma_{1,s}(\omega) \equiv \sigma_{1}(\omega,T\ll T_c)$.  An alternative
method for extracting the superfluid density relies on only the real part of
the dielectric function. Simply put, if upon entering the superconducting state
for $T\ll T_c$ it is assumed that all of the carriers collapse into the
condensate, then $\omega_{pS}\equiv \omega_p$ and $\Gamma \rightarrow 0$, so
that the form of the dielectric function in Eq.~(\ref{eq:drude}) becomes
$\epsilon_1(\omega) = \epsilon_\infty - \omega_{pS}^2/\omega^2$; in the limit
of $\omega \rightarrow 0$, $\rho_s \propto \omega_{pS}^2 = -\omega^2
\epsilon_1(\omega)$.  This is a generic result in response to the formation of
a $\delta(\omega)$ function and is not model dependent.
The value for $-\omega^2\epsilon_1(\omega)$ is shown in Fig.~\ref{fig:sigma2}
for $\Gamma=2\Delta$. There is a small dip near $2\Delta$ (which becomes
somewhat washed out for $\Gamma\gg 2\Delta$), and the curve converges cleanly
in the $\omega \rightarrow 0$ limit. The determination of $\rho_s$ from
$-\omega^2\epsilon_1(\omega)$ has two main advantages: (i) it relies only on
the value of $\epsilon_1(\omega)$ for $T\ll T_c$ and thus probes just the
superfluid response, and (ii) $\rho_s$ is determined in a low-frequency limit,
which removes the uncertainty of the high-frequency cut-off frequency
$\omega_c$ in the FGT sum rule estimates of the condensate. We will distinguish
between values of the condensate determined from $-\omega^2\epsilon_1(\omega)$
as $\rho_s$, and the FGT sum rule as $\omega^2_{pS}$.  The two techniques
should in fact yield the same result, and it is indeed useful to compare the
high-frequency estimates of $\omega^2_{pS}$ with $\rho_s$.

%
%

\begin{figure}[t]
\vspace*{-0.6cm}%
%
%
\centerline{\includegraphics[width=3.8in]{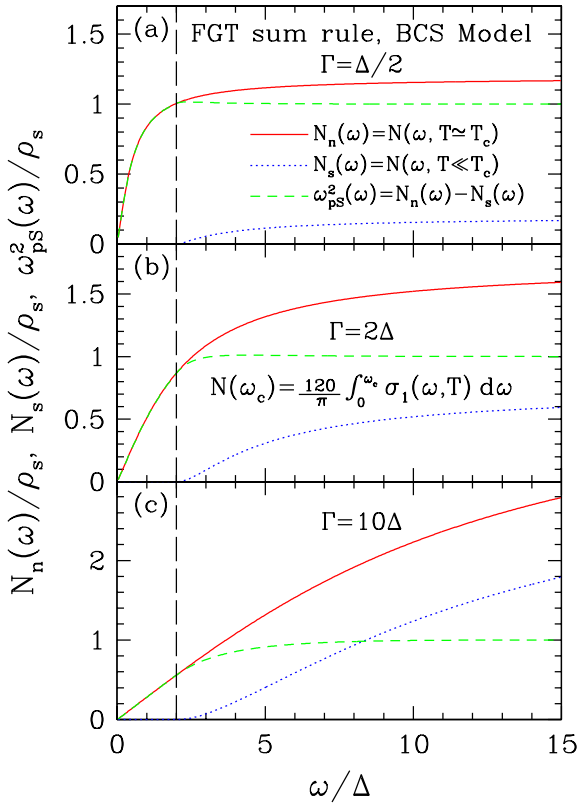}}%
%
%
\vspace*{-1.2cm}%
\caption{The Ferrell-Glover-Tinhkam (FGT) sum rule applied to a BCS model for a
variety of different normal-state scattering rates.  The solid line is the
conductivity sum rule applied to the $\sigma_1(\omega,T\gtrsim T_c)$
[$N_n(\omega)$], while dotted line is for $\sigma_1(\omega,T\ll T_c)$
[$N_s(\omega)$]; the dashed line is the difference between the two.  The curves
have been normalized to the full weight of the condensate $\rho_s$ to yield a
dimensionless ratio. (a) $\Gamma = \Delta/2$, (b) $\Gamma=2\Delta$, and (c)
$\Gamma = 10\Delta$.  The spectral weight transferred from the normal state to
the condensate decreases with increasing $\Gamma$, illustrating the trend from
the clean to dirty-limit case. Even for the largest value of $\Gamma$ chosen, a
large fraction of the condensate is captured by the integral at $\omega_c =
2\Delta$ (long-dashed line).}%
\label{fig:fgt}
\end{figure}

%
%
The rapidity with which the spectral weight of the condensate is captured by
the Ferrell-Glover-Tinkham sum rule is shown in Fig.~\ref{fig:fgt} for three
different choices of the normal-state scattering rate relative to the
superconducting energy gap.  Here the solid line is the conductivity sum rule
applied to $\sigma_1(\omega)$ in the normal state ($T\gtrsim T_c$), effectively
$\omega_p^2$, while the dotted line is the conductivity sum rule for $T\ll
T_c$, which yields $\omega_p^2-\omega_{pS}^2$. The difference between the two
curves is the dashed line, which is simply $\omega_{pS}^2$.  To simplify
matters, in each case the integrals have been normalized with respect to the
strength of the fully-formed condensate $\rho_s$, to yield a dimensionless
ratio.
In Fig.~\ref{fig:fgt}(a) the normal state scattering rate has been chosen to be
$\Gamma=\Delta/2$ (``clean limit'').  It may be observed that nearly all of the
spectral weight in the normal state collapses into the condensate. Furthermore,
the condensate is essentially fully-formed above $2\Delta$.
In Fig.~\ref{fig:fgt}(b) the normal state scattering rate has been chosen to
have an intermediate value $\Gamma=2\Delta$ (the situation depicted in
Fig.~\ref{fig:bcs}). The larger value of $\Gamma$ has the effect of shifting
more of the normal-state spectral weight above $2\Delta$, which reduces the
strength of the condensate.  However, despite the larger value for $\Gamma$ and
the reduced strength of the condensate, it is once again almost fully-formed by
$2\Delta$.
%
%
%
%
Finally, in Fig.~\ref{fig:fgt}(c) a large normal-state scattering rate $\Gamma
= 10\Delta$ is chosen to put the system into the dirty limit.  The large
scattering rate broadens the normal-state conductivity and moves a considerable
amount of the spectral weight to high frequency, thus only a relatively small
amount of the spectral weight is transferred to the condensate.  Despite this,
by $2\Delta$ almost 60\% of the condensate has been captured, and by $4\Delta$
the condensate is almost fully formed.  This demonstrates that with the
exception of the dirty limit, the relevant energy scale for $\omega_c \simeq
2\Delta$.  This result is important to the arguments that are to follow.

%
%
\subsection{YBa$_2$Cu$_3$O$_{6+x}$}
\begin{figure}[t]
\vspace*{-0.8cm}%
%
%
\centerline{\includegraphics[width=3.8in]{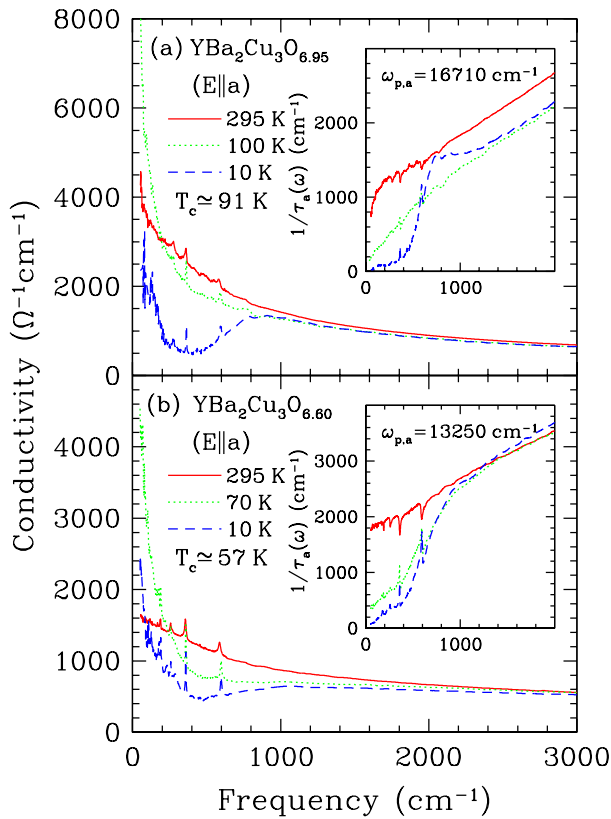}}%
%
%
\vspace*{-1.2cm}%
\caption{The optical conductivity of (a) optimally-doped
YBa$_2$Cu$_3$O$_{6.95}$, and (b) underdoped YBa$_2$Cu$_3$O$_{6.60}$ at room
temperature (solid line), $T\gtrsim T_c$ (dotted line), and $T\ll T_c$ (dashed
line) for light polarized along the {\it a} axis.  The inset in each panel
shows the frequency dependent scattering rate and the estimated value of
$\omega_{p,a}$.  The Drude-like conductivity of the optimally-doped material
narrows somewhat in the normal state but the scattering rate shows no
indication of a pseudogap; below $T_c$ a considerable amount of spectral weight
collapses into the condensate. In the underdoped material, the conductivity
narrows considerably in the normal state and the scattering rate indicates the
opening of a pseudogap; the
condensation is less dramatic than in the optimally-doped case.}%
\label{fig:ybco}
\end{figure}

The temperature dependence of the optical conductivity of optimally-doped
YBa$_2$Cu$_3$O$_{6.95}$ ($T_c\simeq 91$~K) for light polarized along the {\it
a} axis is shown in Fig.~\ref{fig:ybco}(a).  The Drude-like low-frequency
conductivity narrows as the temperature decreases from room temperature to just
above $T_c$; well below the superconducting transition the low-frequency
conductivity has decreased and the missing spectral weight has collapsed into
the condensate.  However, an optical gap is not observed and there is a great
deal of residual conductivity at low frequency.
The conductivity can be reasonably well described using a ``two-component''
model,\cite{tanner92} with a Drude component and a number of bound excitations,
usually Lorentz oscillators.  While the low-frequency conductivity is
satisfactorily described by a Drude term, the midinfrared region is not and a
large number of oscillators are required to reproduce the conductivity.  For
this reason, a generalized form of the Drude model is often adopted where the
scattering rate is allowed to have a frequency dependence\cite{webb86} (in
order to preserve the Kramers-Kronig relation, the effective mass must then
also have a frequency dependence).  The frequency-dependent scattering rate has
the form\cite{puchkov96}
\begin{equation}
  {1\over{\tau(\omega)}} = {{\omega_p^2}\over{4\pi}} \mathrm{Re} \left[ {1\over
  {\tilde\sigma(\omega)}} \right].
\label{eq:tau}
\end{equation}
The value for the plasma frequency used to scale the expression in
Eq.~(\ref{eq:tau}) has been estimated using the conductivity sum rule for
$\sigma_{1,a}(\omega, T\simeq T_c)$, using $\omega_c\simeq 1$~eV, which yields
a value for $\omega_{p,a}\simeq 16700$~cm$^{-1}$, or about 2~eV
(Ref.~\onlinecite{homes99}).  The frequency-dependent scattering rate
$1/\tau_a(\omega)$ is shown in the inset of Fig.~\ref{fig:ybco}(a), and in the
normal state shows a monotonic increase with frequency, and an overall downward
shift with decreasing temperature.  Below $T_c$, there is a strong suppression
of $1/\tau_a(\omega)$ at low frequency associated with the formation of the
superconducting gap, with a slight overshoot and then the recovery of the
normal-state value at high-frequency.\cite{basov02,abanov02} This behavior is
characteristic of optimally-doped and overdoped materials.

%
%

%
%
The behavior of the oxygen-underdoped material YBa$_2$Cu$_3$O$_{6.60}$
($T_c\simeq 57$~K) for light polarized along the {\it a} axis, shown in
Fig.~\ref{fig:ybco}(b), shows some significant differences from the
optimally-doped material.  The Drude-like conductivity at room temperature is
extremely broad.  However, at $T\simeq T_c$ the Drude-like conductivity has
narrowed dramatically, and there has been a significant shift of spectral
weight to low frequencies. For $T\ll T_c$ the low frequency conductivity has
decreased, indicating the formation of a condensate.  However, the effect is
not as dramatic as it was in the optimally-doped material, indicating that the
strength of the condensate is not as great.  Once again, there is a
considerable amount of residual conductivity at low frequency for $T\ll T_c$.
The frequency-dependent scattering rate is shown in the inset, along with the
estimated value of $\omega_{p,a}\simeq 13\,250$~cm$^{-1}$, estimated from the
conductivity sum rule.  As the temperature decreases in the normal state
$1/\tau_a(\omega)$ decreases rapidly at low frequency, which is taken to be
evidence for the formation of a pseudogap.\cite{puchkov96,timusk99} The large
drop in the normal-state scattering rate for $1/\tau_a(\omega\rightarrow 0)$ is
a reflection of the dramatic narrowing of the conductivity and is an indication
that the reduced doping has not created a large amount of scattering due to
disorder --- on this basis, the system is {\it not} in the dirty limit.

%
%
\begin{figure}[t]
\vspace*{-0.4cm}%
%
%
\centerline{\includegraphics[width=3.8in]{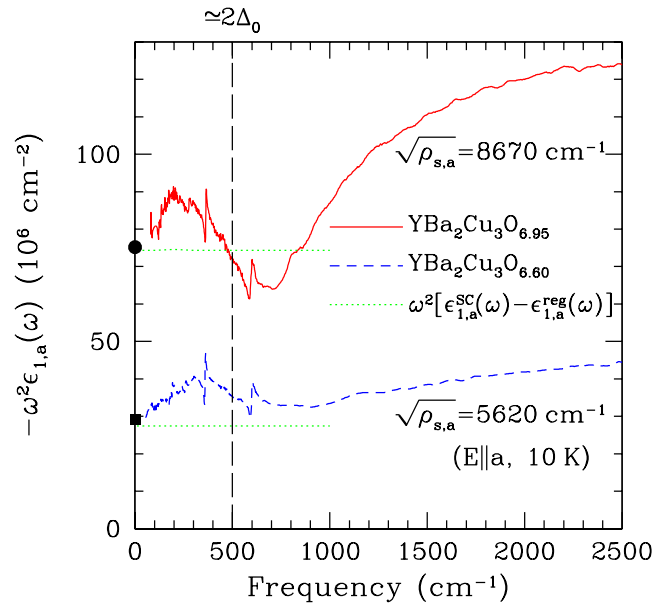}}%
%
%
\vspace*{-1.2cm}%
\caption{The function $-\omega^2\epsilon_{1,a}(\omega)$ vs frequency for
optimally doped YBa$_2$Cu$_3$O$_{6.95}$ (solid line) and underdoped
YBa$_2$Cu$_3$O$_{6.60}$ (dashed line) along the {\it a} axis at $\simeq 10$~K
($T\ll T_c$).  The superfluid density is
$\rho_s=-\omega^2\epsilon_{1,a}(\omega\rightarrow 0)$; taking the squares to
render the units the same as those of a plasma frequency yields
$\sqrt{\rho_{s,a}}=8670\pm 90$~cm$^{-1}$ and $\sqrt{\rho_{s,a}}=5620\pm 60
$~cm$^{-1}$ for the optimally and underdoped materials, respectively.  Note
also that in both materials there is a slight suppression of
$-\omega^2\epsilon_{1,a}(\omega)$ in
the $500-700$~cm$^{-1}$ region, close to the estimated value of $2\Delta_0$.}%
\label{fig:rho}
\end{figure}

%
%
The superfluid density $\rho_{s,a}$ has been estimated from the response of
$-\omega^2\epsilon_{1,a}(\omega)$ in the zero-frequency limit for $T\ll T_c$
for the optimally and underdoped materials, shown in Fig.~\ref{fig:rho}.  The
estimate of $\rho_{s,a}$ assumes that the response of $\epsilon_{1,a}(\omega)$
at low frequency is dominated by the condensate, but it has been shown that
along the {\it c} axis, there is enough residual conductivity to affect
$\epsilon_{1,c}(\omega)$ and thus the values of $\rho_{s,c}$, typically
resulting in an overestimate of the strength of the condensate.\cite{homes98,
dordevic02} The presence of residual conductivity for $T\ll T_c$ suggests that
$\rho_{s,a}$ may be overestimated in this case as well.  However, as we noted
earlier, the real part of the conductivity in the superconducting state may be
expressed as a regular part due to unpaired carriers, and a $\delta(\omega)$
function at zero frequency; the response of $\epsilon_{2,a}^{SC}(\omega)$ is
limited to the $\delta(\omega)$ function, which is zero elsewhere. However,
$\epsilon_{2,a}(\omega)$ has been determined experimentally to be non-zero: if
we refer to this as $\epsilon_{2,a}^{reg}(\omega)$, then $\epsilon_{1,a}^{reg}
(\omega)$ may be determined through the Kramers-Kronig relation, and the
superfluid density estimated as\cite{dordevic02}
\begin{equation}
  \rho_{s,a}(\omega)=\omega^2 \left[ \epsilon_{1,a}^{SC}(\omega)-
                     \epsilon_{1,a}^{reg}(\omega) \right],
\end{equation}
which should be a constant.  This is shown in Fig.~\ref{fig:rho} at low
frequency as the dotted lines.
This method of estimating $\rho_{s,a}$ agrees well with the extrapolated values
of $-\omega^2\epsilon_1(\omega)$ in the $\omega\rightarrow 0$ limit, and
indicates that if there is a correction to $\rho_{s,a}$ associated with the
residual conductivity for $T\ll T_c$, then it is quite small. The estimated
values for the condensate are $\sqrt{\rho_{s,a}} = 8670\pm 90$~cm$^{-1}$ and
$5620\pm 60$~cm$^{-1}$ for the optimally and underdoped materials,
respectively; these estimates are in good agreement with previous
values.\cite{homes99} In both materials there is a slight suppression of
$-\omega^2\epsilon_1(\omega)$ in the $500-700$~cm$^{-1}$ region, which is in
agreement with estimates for the superconducting gap maximum $2\Delta_0\simeq
500$~cm$^{-1}$ (adopting the notation for a {\it d}-wave superconductor) in
overdoped YBa$_2$Cu$_3$O$_{6.99}$ (Ref.~\onlinecite{lu01}). Studies of other
cuprate systems suggest that the gap maximum increases with decreasing
doping,\cite{harris96,miyakawa99,anderson00} despite the reduction of $T_c$.

%
%
\begin{figure}[t]
\vspace*{-0.6cm}%
%
%
\centerline{\includegraphics[width=3.8in]{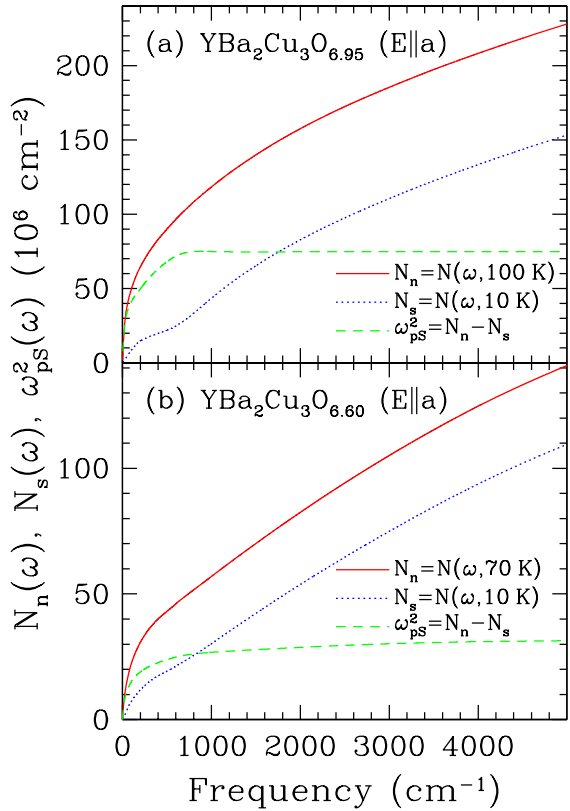}}%
%
%
\vspace*{-1.2cm}%
\caption{The conductivity sum rules applied to (a) optimally-doped
YBa$_2$Cu$_3$O$_{6.95}$ and (b) underdoped YBa$_2$Cu$_3$O$_{6.95}$ for light
polarized along the {\it a} axis for $T \gtrsim T_c$ [$N(\omega)$, solid line]
and for $T\ll T_c$ [$N_s(\omega)$, dotted line]; the difference is the estimate
of the strength of the condensate from the Ferrell-Glover-Tinkham sum rule
(dashed line).  The condensate saturates in the upper panel by $\simeq 800$
cm$^{-1}$, while in the lower panel the frequency at which the full weight of
the condensate is recovered seems to be much higher.  In the upper panel the
magnitudes of the curves are greater than in the lower panel; a reflection of
the decreased carrier concentration.}%
\label{fig:sums}
\end{figure}

%
%
The integrated values of the conductivity in the normal ($T\gtrsim T_c$) and
superconducting ($T\ll T_c$) states are indicated by the solid [$N_n(\omega)$]
and dashed [$N_s(\omega)$] lines for YBa$_2$Cu$_3$O$_{6.95}$ and
YBa$_2$Cu$_3$O$_{6.60}$ along the {\it a} axis in the upper and lower panels of
Fig.~\ref{fig:sums}, respectively.
For YBa$_2$Cu$_3$O$_{6.95}$, $N_n(\omega)$ increases rapidly with frequency,
but does not display any unusual structure.  On the other hand, $N_s(\omega)$
evolves more slowly, and has several inflection points at low frequency which
are thought to be related to the peaks in the electron-boson spectral
function.\cite{tu02} The difference between the two curves $\omega_{pS}^2 =
N_n(\omega)-N_s(\omega)$ is shown by the dashed line in Fig.~\ref{fig:sums}(a).
This quantity increases quickly and then saturates above $\approx
800$~cm$^{-1}$ to a constant value. this plot is reminiscent of the BCS
material with moderate scattering, discussed in Fig.~\ref{fig:fgt}(b).
The sum rules applied to YBa$_2$Cu$_3$O$_{6.60}$ shown in
Fig.~\ref{fig:sums}(b) are similar to the optimally-doped case.  However, the
overall magnitude has decreased, a reflection of the decreased carrier
concentration within the copper-oxygen planes in the underdoped material. While
the condensate is also lower, it now appears that it does not saturate as
quickly as was the case in the optimally-doped material.

%
%
\begin{figure}[t]
\vspace*{-0.5cm}%
%
%
\centerline{\includegraphics[width=3.8in]{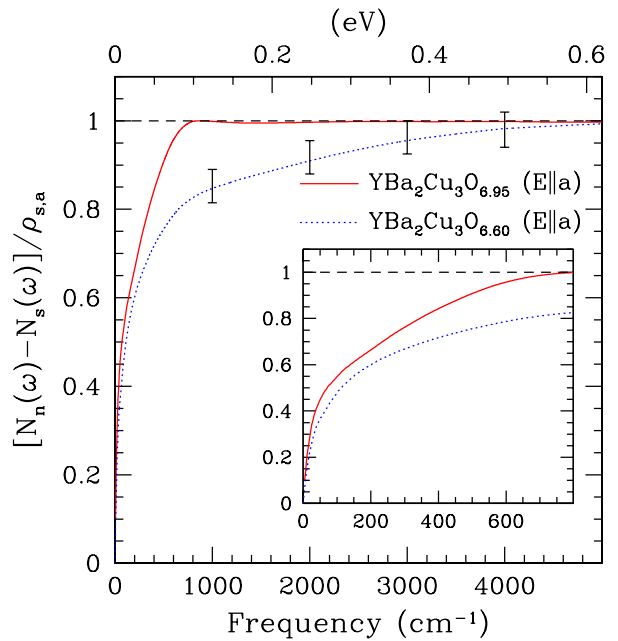}}%
%
%
\vspace*{-0.8cm}%
\caption{The normalized weight of the condensate $[N_n(\omega)-N_s(\omega)] /
\rho_{s,a}$ for optimally-doped YBa$_2$Cu$_3$O$_{6.95}$ (solid line) and
underdoped YBa$_2$Cu$_3$O$_{6.60}$ (dotted line) along the {\it a} axis
direction.  The curves describing the condensate have been normalized to the
values of $\rho_{s,a}$ shown in Fig.~\ref{fig:rho}.  The condensate for the
optimally-doped material has saturated by $\simeq 800$ cm$^{-1}$, while in the
underdoped material the condensate is roughly 80\% formed by this frequency,
but the other 20\% is not recovered until much higher
frequencies.  The error bars on the curve for the underdoped material indicate
the uncertainty associated with the FGT sum rule.  Inset: The low-frequency
region.}%
\label{fig:weight}
\end{figure}

%
%
A more detailed examination of the evolution of the weight of the condensate
for YBa$_2$Cu$_3$O$_{6.95}$ (solid line) and YBa$_2$Cu$_3$O$_{6.60}$ (dotted
line), normalized to the values of $\rho_{s,a}$ determined in
Fig.~\ref{fig:rho}, is shown in Fig.~\ref{fig:weight}.  The error associated
with the FGT sum rule has been determined in the following way.  The optical
conductivity has been calculated for $R(\omega, T)\pm 0.1$\% for $T\gtrsim T_c$
and $T\ll T_c$, the normal and superconducting states, respectively.  The FGT
sum rule is then applied to the resulting high and low values for the
conductivity, and the error limits are taken as the deviation from the curve
generated simply from $R(\omega ,T)$, which are estimated to be about $\pm
3$\%.  It should be noted that most of the uncertainty is introduced when the
reflectance is close to unity, as $\sigma_1 \propto 1/(1-R)$ and even small
uncertainties in the reflectance can lead to large errors in the optical
conductivity.
When the FGT sum rule is exhausted, the ratio is unity by definition.  For the
optimally-doped material, this occurs rapidly and 90\% of the spectral weight
in the has been recovered by about 500~cm$^{-1}$, and the ratio approaches
unity at $\omega_c \approx 800$~cm$^{-1}$, and remains constant even out to
very high frequencies (over 0.5~eV). This rapid formation of the condensate has
also been observed in the optimally-doped materials\cite{tanner98,tuprivate}
La$_{1.85}$Sr$_{0.15}$CuO$_4$ ($\omega_c \simeq 0.05$~eV),
Bi$_2$Sr$_2$CaCu$_2$O$_{8+\delta}$ ($\omega_c \simeq 0.1$~eV), as well as in
the electron-doped material\cite{homes97} Nd$_{1.85}$Ce$_{0.15}$CuO$_{4+
\delta}$ ($\omega_c \simeq 0.06$~eV).  In all of these cases the integral
saturates and is constant to over 0.5~eV.
In contrast, only about 80\% the spectral weight in underdoped
YBa$_2$Cu$_3$O$_{6.60}$ has formed by 800 cm$^{-1}$; the 90\% threshold is not
reached until $\simeq 1800$~cm$^{-1}$ and the remaining spectral weight is
recovered only at much higher frequencies ($\omega_c \gtrsim 5000$~cm$^{-1}$).
%
%
In the case of the underdoped material, the plot has only been shown to the
point where the FGT sum rule is recovered.  If the plot is extended to $\sim
1$~eV, then the integral will increase to a value about $\approx 3$\% over
unity, which is within the estimated error for the FGT sum rule.  This slow
increase above unity may be an indication of one of two things: (i) $\omega_c$
may be larger than has been previously estimated, which would be consistent
with estimates of $\omega_c\simeq 2$~eV in the underdoped
Bi$_2$Sr$_2$CaCu$_2$O$_{8+\delta}$ materials,\cite{syro01} or (ii) it is
possible that when the sum rule is extended to high frequencies (i.e. of the
order of eV) it may be incorporating temperature-dependent bound excitations.
However, the absence of this behavior in the optimally-doped system suggests
that such an excitation is restricted to the underdoped materials.

%
%
It is tempting to draw an analogy with the BCS dirty-limit case and argue that
the spectral weight in the underdoped material has been pushed to higher
frequency in response to an increase in the normal-state scattering rate.
However, there are two important points that argue against this interpretation.
First, an examination of scattering rate in the insets of Fig.~\ref{fig:ybco}
for $T\gtrsim T_c$ indicates that the $1/\tau_a(\omega\rightarrow 0) \lesssim
200$~cm$^{-1}$ for both materials.
Second, if the conductivity is fitted using a two-component Drude-Lorentz
model, then the nature of the low-frequency conductivity places hard
constraints on the width of the Drude peak;\cite{quijada99} for $T\gtrsim T_c$
then $\Gamma\simeq 140$~cm$^{-1}$ for optimally-doped YBa$_2$Cu$_3$O$_{6.95}$,
and $\Gamma\simeq 100$~cm$^{-1}$ for underdoped YBa$_2$Cu$_3$O$_{6.60}$.  In
each case, $\Gamma < 2\Delta_0$, indicating that while $\Gamma$ may have an
unusual temperature dependence, close to $T_c$ these materials are not in the
dirty limit.  Thus, the larger energy scale in the underdoped system has a
different origin.
While none of the cuprate superconductors are truly good metals, it has been
suggested that for $T\gtrsim T_c$ the overdoped materials may resemble a Fermi
liquid.\cite{proust02} The rapid convergence of $\rho_{s,a}$ in the
optimally-doped material is what would be expected in a BCS system in which the
normal state is a Fermi liquid.  On the other hand, it is recognized that the
underdoped materials are bad metals\cite{emery95b} and exhibit non-Fermi liquid
behavior, and $\omega_c \gg 2\Delta_0$ is required to recover the FGT sum rule.
The two types of behavior observed in the optimal and underdoped materials
suggests that the nature of the electronic correlations in the normal state
play a role in determining the different aspects of the superconductivity
observed in these materials.\cite{norman02}

%
%
\subsection{Kinetic energy and the sum rule}
The unconventional nature of the superconductivity in the cuprate systems has
lead to the suggestion that the condensation may be driven by changes in the
kinetic rather than the potential energy.\cite{hirsch92} In such a case the FGT
sum rule for the in-plane conductivity must be modified to take on the
form\cite{hirsch00a}
\begin{eqnarray}
  \rho_s & = & {120\over\pi} \int_{0+}^{\omega_c} \left[
               \sigma_{1,n}(\omega) - \sigma_{1,s}(\omega) \right]\,d\omega \nonumber \\
         &   & + {{e^2a^2}\over{\pi c^2\hbar^2}}
    \left[ \langle -T_s \rangle - \langle -T_n \rangle\right],  \\
    & \equiv & \delta{A}_l + \delta{A}_h, \nonumber \\
    \label{eq:kinetic} \nonumber
\end{eqnarray}
where $a$ is the lattice spacing, and $T$ is that part of the in-plane kinetic
energy associated with the valence band,\cite{hirsch00a, hirsch00b} and the
subscripts {\it n} and {\it s} refer to $T\simeq T_c$ and $T\ll T_c$,
respectively. (The expression has a slightly different form for the {\it c}
axis.\cite{chakravarty98,chakravarty99,munzar01})  The low-frequency term
$\delta{A}_l$ is simply the FGT sum rule, while $\delta{A}_h$ corresponds to
the high-frequency part of the integral, which is in fact the kinetic energy
contribution.
%
%
In a system where the kinetic energy plays a prominent role the FGT sum rule
may appear to be violated.  However, the maximum condensation energy for
YBa$_2$Cu$_3$O$_{6.95}$ based on specific heat
measurements\cite{loram01,vdmarel02} is about 0.2~meV per (in-plane) copper
atom. Assuming that the condensation energy is due entirely to the changes in
the in-plane kinetic energy, this yields $\delta{A}_h\approx 2\times
10^5$~cm$^{-2}$, which represents less than 0.3\% of the spectral weight of the
condensate.\cite{energy}  Because of the limited accuracy of the FGT sum rule
no statement may be made regarding changes in the in-plane kinetic energy.
However, the general observation that the in-plane sum rule is preserved may
have consequences for the {\it c} axis.
%

%
%
\subsection{Sum rules along the \boldmath $c$ \unboldmath axis}
The optical properties of YBa$_2$Cu$_3$O$_{6+x}$ have been examined in some
detail along the poorly-conducting {\it c} axis.\cite{homes93b,schutzmann94,
homes95a,tajima97}  The optical conductivity (especially in the underdoped
materials) is dominated by the unscreened phonons.\cite{homes95b,schutzmann95}
Given that large changes in the phonon spectrum have been observed at low
temperature in the underdoped materials,\cite{homes95b,schutzmann95,
munzar99,gruninger00,bernhard00} the application of the FGT sum rule must be
treated with some care; in the studies cited here,\cite{basov99,basov01} the
integral has been truncated at $\omega_c \simeq 800$~cm$^{-1}$ ($0.1$~eV).
The application of the FGT sum rule along the {\it c} axis has shown that while
this sum rule is not violated in the optimally-doped materials, as these
materials become increasingly underdoped and a pseudogap has formed, the FGT
sum rule is violated to varying degrees, with more than 50\% of the {\it
c}-axis spectral weight is missing at low temperature.\cite{basov99,basov01,
kuzmenko03} The violation of the FGT sum rule has been proposed as evidence for
a kinetic energy contribution,\cite{chakravarty98,chakravarty99} although there
have also been other interpretations of this phenomena.\cite{ioffe99,munzar01}
The implication is that the missing spectral weight is recovered at high
frequency, but it is unclear at precisely what point this occurs.
%
%
%
%
%
%
In underdoped material, if the value for $\omega_c=800$~cm$^{-1}$ in the
optimally-doped materials is used, then the in-plane FGT sum rule will appear
to be violated. However, it has been shown that extending the integral from
$\omega_c \simeq 800$ to $\gtrsim 5000$ cm$^{-1}$ results in the recovery of
the in-plane sum rule.  We speculate that if the cut-off frequency for the FGT
sum rule along the {\it c} axis is increased to the same value where the
in-plane sum rule was recovered ($\omega_c\gtrsim 5000$~cm$^{-1}$) then the
spectral weight would be recovered and the {\it c} axis sum rule would yield
$\rho_{s,c}$. However, the general consensus at this time is that the
conductivity data is not yet sufficiently precise along the {\it c} axis to
confirm this prediction, so this remains a subject of some debate.

%
%
\section{Conclusions}
In the BCS model the relevant energy scale to recover the strength of the
condensate $\rho_s$ is the superconducting energy gap ($\omega_c \simeq2
\Delta$), slightly larger in the dirty-limit case.
Conductivity and FGT sum rules have been examined for light polarized along the
{\it a} axis direction in the optimally doped YBa$_2$Cu$_3$O$_{6.95}$ and
underdoped YBa$_2$Cu$_3$O$_{6.60}$ high-temperature superconductors to study
the evolution of the spectral weight in these materials.  Within the
sensitivity of the experiment the FGT sum rule is obeyed in both materials. The
energy scale required to recover the full strength of the condensate in the
optimally doped material is $\omega_c\simeq 800$~cm$^{-1}$ ($\simeq
2\Delta_0$), in good agreement with the predicted behavior of the BCS model.
However, the energy scale in the underdoped materials is much higher,
$\omega_c\gtrsim 5000$~cm$^{-1}$. This effect can not be attributed to
dirty-limit effects in response to increased normal-state scattering, since for
$T\simeq T_c$, $\Gamma < 2\Delta_0$ in both materials.
The two types of behavior above and below $T_c$ observed in the optimal and
underdoped materials suggests that the nature of the electronic correlations in
the normal state determine the different aspects superconductivity in these
materials,\cite{norman02} and the degree to which the kinetic energy may play a
role.

%
%
\begin{acknowledgments}
We are grateful to D. van der Marel for insights regarding the specific heat
results. We would also like to thank D. N. Basov, V. J. Emery, A. Chubukov, S.
A. Kivelson, F. Marsiglio, C. Pepin, M. Strongin, D. B. Tanner, T. Timusk, J.
M. Tranquada, and J. J. Tu for useful discussions. This work was supported by
the Department of Energy under Contract No. DE-AC02-98CH10886 and by the
Canadian Institute for Advanced Research.
\end{acknowledgments}

%
%
\bibliography{sumrule}

\begin{thebibliography}{67}
\bibitem[{nef()}]{neff}
\bibinfo{note}{The term $120/\pi$ before the integral assumes that the units of
  conductivity are in $\Omega^{-1}$cm$^{-1}$, and that the frequency is in
  cm$^{-1}$, so that the integral yields cm$^{-2}$. The factor in front of the
  integral is sometimes expressed as $2m^\ast V_c/\pi e^2$, where $V_c$ is the
  volume of the unit cell, in which case the integral yields the effective
  number of carriers.}
\end{thebibliography}

\end{document}